# Ion-Beam-Induced Defects in CMOS Technology: Methods of Study

Yanina G. Fedorenko

Additional information is available at the end of the chapter

**Abstract**

Ion implantation is a nonequilibrium doping technique, which introduces impurity atoms into a solid regardless of thermodynamic considerations. The formation of metastable alloys above the solubility limit, minimized contribution of lateral diffusion processes in device fabrication, and possibility to reach high concentrations of doping impurities can be considered as distinct advantages of ion implantation. Due to excellent controllability, uniformity, and the dose insensitive relative accuracy ion implantation has grown to be the principal doping technology used in the manufacturing of integrated circuits. Originally developed from particle accelerator technology, ion implanters operate in the energy range from tens eV to several MeV (corresponding to a few nms to several microns in depth range). Ion implantation introduces point defects in solids. Very minute concentrations of defects and impurities in semiconductors drastically alter their electrical and optical properties. This chapter presents methods of defect spectroscopy to study the defect origin and characterize the defect density of states in thin film and semiconductor interfaces. The methods considered are positron annihilation spectroscopy, electron spin resonance, and approaches for electrical characterization of semiconductor devices.

## 1. Introduction

Applications of ion implantation require an understanding of the lattice defects, which largely control the optical and electrical properties of semiconductors. Characterization techniques such as secondary ion mass spectrometry, spreading resistance, carrier and mobility profiling,

Rutherford backscattering, ion channeling, and transmission electron microscopy with examples of using these techniques to investigate the dopant distribution in the implanted samples, characterize dopants that are electrically active, examine accumulation of the ion beam induced defects, and resolve their structure have been reviewed in the literature [1]. As the main feature of ion implantation is the formation of point defects in the energetic ion collisions, it is natural to present additional methods employed in semiconductor research to study atomic origin and electrical activity of technologically relevant imperfections. The prime attention will be given to characterization techniques invented in technological development of the Si/SiO$_2$ system, though examples of other materials systems, which can be studied by application of positron annihilation spectroscopy, electron spin resonance spectroscopy, and (photo) electrical methods are provided.

## 2. Positron annihilation spectroscopy

Positron annihilation spectroscopy (PAS) is now a well-established tool to characterize electronic and defect properties of bulk solids, thin films, and surfaces. PAS allows studying the electronic structure of defects in solids. The imperfections are represented by small volume defects such as vacancies, vacancy clusters, and free volume defects. Positron beams can be applied to study defects in metals, semiconductors, composite materials, and thin film systems of different crystalline structure and chemical bonding. Methodologically, PAS mainly considers the three experimentally accessible dependences schematically indicated in **Figure 1**: (i) the time-dependent distribution of annihilating photons; (ii) the angular distribution of annihilating photons; and (iii) the Doppler broadening of the 0.511 MeV annihilation line. While the time-dependent distribution of photons bears information on the electron density in the vicinity of the annihilation event, the latter two photon characteristics provide information on the electron momentum distribution. The positron lifetime gives more integral information than the momentum measurements regarding the region from which the positron annihilates. In the case of a defect-containing sample, the average electron density at a defect site can be rather defect-specific. This suggests position lifetime measurements are suitable for investigating vacancy-clustering processes in rapidly quenched or (ion) irradiated materials. The momentum measurements can also yield detailed defect-specific information. The positron energy may vary allowing examination of the depth distribution of defects in solids and interfaces. Other direct experimental methods including transmission electron microscopy and atomic diffusion are less capable in detecting open volume defects located at interfaces and surfaces. The threshold defects concentration ensured by PAS is $10^{14}$ to $10^{15}$cm$^{-3}$.

The physics of positron annihilation spectroscopy has been explained in textbooks [2, 3] and research articles [4, 5]. A positron injected into a solid becomes thermalized within a few picoseconds by ionizing collisions, plasmon and electron-hole excitations, and phonon interactions. If lattice defects are present in the material, the positron can be trapped by these imperfections. Lattice imperfections (vacancies, vacancy clusters, or dislocations), open volumes, nanoclusters, and the surface states can serve as potential wells, which effectively trap positrons. Within hundreds of picoseconds, a positron in a solid annihilates with an electron yielding two gamma rays.

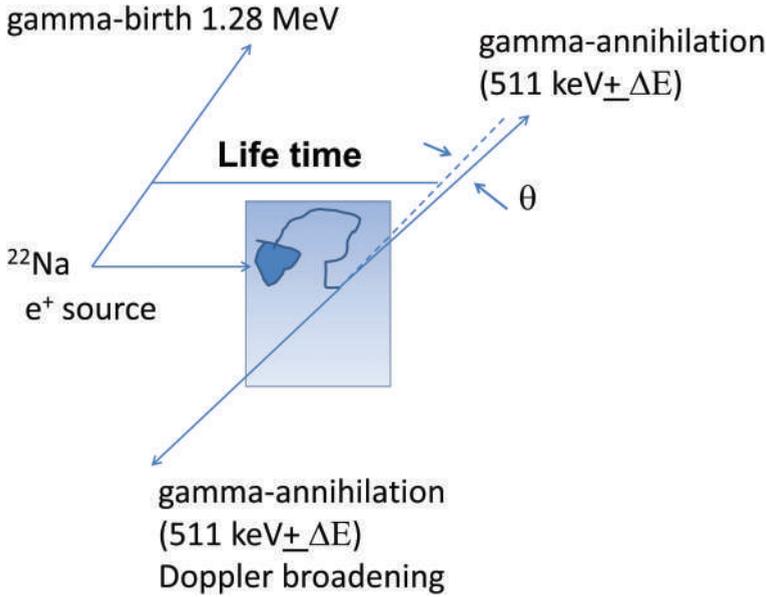

**Figure 1.** Schematic representation of positron annihilation indicating the basis for the three experimental techniques of positron annihilation spectroscopy: lifetime, angular correlation, and Doppler broadening.

The distribution of the $\Delta t$ values for a number of these events, measured in a PAS lifetime experiment, provides the total electron density in the region of positron-electron annihilation. The positron annihilation rate $\lambda$ is the reciprocal of the positron lifetime and can be described by the overlap integral of the electron $\rho^-(r)$ and positron $\rho^+(r)$ densities [4]:

$$\lambda = \pi r_0^2 c \iiint \rho^-(r)\rho^+(r) d^3 r, \tag{1}$$

where $r_0$ is the classical electron radius and c is the velocity of light.

Because energy and momentum are conserved in the annihilation process, the two gamma rays resulting from the electron-positron pair annihilation each have energy equal to the rest-mass energy of an electron or positron ($mc^2$ = 511 keV) and ± an energy increment $\Delta E$; the two gamma rays propagate in opposite directions with some deviation $\theta$. Since the thermal energies of the positions are about $kT$, the values of $\Delta E$ and $\theta$ correspond only to the momenta of the annihilating crystal electrons. The similarity of information available from Doppler-broadening spectra $P(\Delta E)$ and angular-correlation curves $N(\theta)$ can be inferred by comparing the expressions for $N(\theta)$ and $P(\Delta E)$ in terms of the independent-particle-model (IPM) probability, $R(p)$, that positron-electron annihilation yields 2$\gamma$-emission with total momentum $p$:

$$R(p) = \pi r_0^2 c \sum_k n_k |\iiint e^{-ipr} \Psi_+(r)\Psi_-(r)|^2 d^3 r, \tag{2}$$

where $\Psi_+(r)$ and $\Psi_-(r)$ are the positron and electron wave functions, respectively, $n_k$ is the Fermi function, and $k$ represents both the electron wave vector $k$ and the band index. The expression for $N(\theta)$ and $P(\Delta E)$ is represented as $N(\theta_z) = \iint R(\boldsymbol{p}) dp_x dp_y$ and $P(\Delta E_x) = \iint R(\boldsymbol{p}) dp_y dp_z$. The IPM approximation ignores the effects of positron-electron correlations in the solid assuming the particles act independently. The treatment of the electron-positron correlation, i.e., the enhancement of the electron density at a positron trapped by a defect site has been considered in Ref. [6]. The theory developed in this work considers the two-particle representation of an annihilating positron-electron pair. The IPM approximation is used to calculate the momentum distribution for each electron state. The individual contributions are weighted by the corresponding partial annihilation rates. The partial rates are calculated within the generalized gradient approximation. This approach was found useful when considering the momentum region where the uppermost core electron states dominate. The analysis of the momentum distribution curves up to rather large momenta becomes possible enabling identification of the chemical environment where the annihilation event has occurred. The one-dimensional momentum distribution of the annihilating electron-positron pairs can be extracted from the measurement of the Doppler broadening of the annihilation radiation. Generally, the positron-enhanced electron density can be accounted for if a constant, multiplicative factor (the enhancement factor) is used to take the many-body effects into account, although different enhancement factors must be used for valence and core electrons consistent with their degree of tight-binding.

A typical positron lifetime experiment has been described in work [7]. It can be performed by using a radioactive $^{22}$Na as a positron source. The positron source material can be deposited on a sample or sealed in foil, then placed between two identical samples under study. The decaying Na nuclei emit a high energy photon at 1.2745 MeV, which is used as a start signal for the positron lifetime measurement, while a stop signal is characterized by 511 keV photons. The photons serving as start and stop signals are detected by scintillating detectors coupled with photomultiplier tubes. Detectors are chosen to optimize scintillating efficiency and resolution. The use of digitization of the detector pulses significantly simplifies the postmeasurement signal analysis. The measured positron lifetime spectrum is exponential and reveals several features such as the background noise, the time resolution, and annihilations in the source. The background noise is determined by the source activity and arises due to rapid emissions of positrons that produce false coincidences. Further, the data analysis methods are also described in Ref. [7]. Except for the least-squares fitting of the positron life time spectrum, the inverse Laplace transform and the Bayesian-probability methods have been developed. The latter two methods do not require the number of lifetime components to be *a priori* fixed and can be used if continuous lifetime distributions are expected.

The surroundings of the vacancy defect can be studied with coincidence Doppler broadening spectroscopy measurements. Nonzero electron and positron momentum causes a Doppler shift of the annihilation photons. The Doppler shift is determined by the momentum of electrons since positrons in a solid are thermalized. Analysis of the Doppler broadening of annihilation radiation provides a sensitive method of defect characterization by extracting the momentum distribution of the electrons. It allows examining high-momentum core electrons. The principle of the method lies in the analysis of the positron annihilation line shape, which directly corresponds to the distribution of momentum of electron-positron pairs as shown in **Figure 2**. The momentum itself is measured from the amount of the Doppler shift of the emitted photons. In the coincidence Doppler broadening spectroscopy developed in works [8, 9] determination of energy of both γ rays is done simultaneously. Coincidence measurements of annihilation photons reduce the background signal by several orders of magnitude and allow observation of the high-momentum part of the spectrum, which stems from positrons annihilated by core electrons making possible identification of chemical elements surrounding a positron annihilation site.

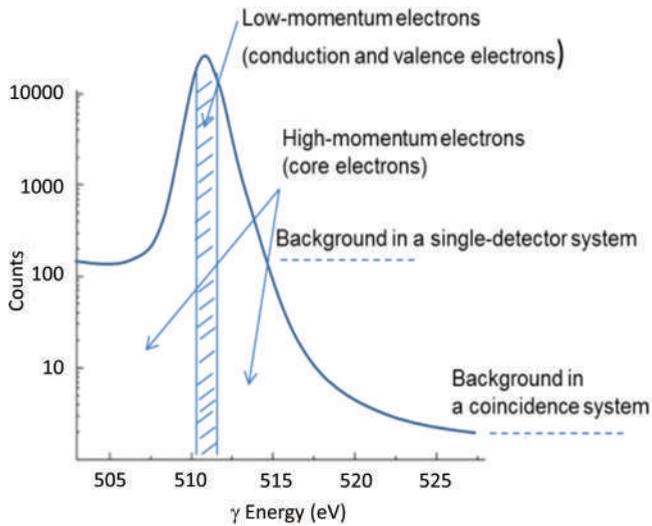

**Figure 2.** A typical annihilation line. After Ref. [144].

The discovery of slow positron emitters enabled analysis of solid surfaces [10, 11]. Slow positron beams are utilized for nondestructive depth profiling of defects in surfaces and interfaces, low energy positron diffraction, and positron remission microscopy studies on surfaces. A moderator single crystal metal film (Au, Cu, W, Ta) was used to produce slow positron beams [12]. The thermalized slow positrons are emitted from the metal surface spontaneously owing to the negative positron-surface affinity. Since slow positron beam generation is a surface process, it is sensitive to surface contaminations such as carbon, oxygen, and the surface defects [13, 14]. Energy loss mechanisms and the positron processes in condensed matter are described in Ref. [15]. Except being ejected from the surface, positrons can form a positronium ($P_s$) by capturing a surface electron. This bound state decays from either a singlet state, p-$P_s$ ($^1S_0$) or a triplet state, o-$P_s$ ($^3S_1$), each having unique annihilation characteristics [16]. Positrons can become trapped by the surface states or reflected back to the interior from the surface.

When a slow positron annihilates with a core electron, the released energy can be transferred to another electron, which can be ejected and detected out of the surface. Weiss et al. [17] were first to demonstrate that a low-energy positron creates core holes through matter-antimatter annihilation generating Auger electrons with high efficiency and extremely low secondary electron background. The latter is feasible to obtain by using incident beam energy below the secondary electron emission threshold. Positron-annihilation-induced Auger-electron spectroscopy (PAES) is advantageous due to increased surface selectivity in systems where the localization of the positron at the surface causes the excitation volume to be restricted almost to the top atomic layer. In addition, because calculated PAES intensities are very sensitive to the spatial extent of the positron wave function, PAES measurements provide an important test for models of the positron surface state. This technique has been proved to be a useful tool for determining surface composition, thin film and nanocrystal characterization, and surface diffusion of atoms.

Also, positrons can be used in diffraction experiments having the advantage that interaction with solids can be easier modeled due to the sign of the scattering potential (the scattering potential between the positron and the atomic nucleus is repulsive) and the total reflection, which is only present in the positron diffraction [18, 19]. The interaction of an energetic positron with the solid may differ from that of electrons of the same energy. The differences can be associated with the relative differential and total elastic cross sections and also with the different energy loss processes for the two particles in a solid. At low energy, the inelastic mean-free path of a positron is shorter than that of an electron leading to an increased surface sensitivity of positrons. This is especially useful in examining the features of reconstructed surfaces, adsorbates, single adsorbed layers and their spacing to the substrate as well as layers with a nominal thickness in the submonolayer range. The positron scattering cross sections are marginally dependent on the specific element enabling analysis of compounds comprising unlike atoms.

High energy diffraction of positrons generates two-dimensional (2D) pattern similar to electrons, although there are several differences due to differences in the ion-core interaction and crystal potential between positrons and electrons. Kikuchi lines stemming from multiple-scattering of electrons are not observed in diffraction of high-energy positrons. The most notable feature is in the total reflection of positrons at surfaces. The positron diffraction near the critical angle is especially sensitive to the topmost atomic surface layer whereas at the critical angle for total reflection in X-ray diffraction, which is usually less than 0.2° the penetration depth of the photons into the solid still amounts to a few nanometers. Surface sensitivity of positron techniques is especially suited to near-surface measurements, which are particularly relevant to ion beam modified devices.

The technique of positron annihilation spectroscopy in conjunction with a slow positron beam has been proposed for the monitoring of ion implantation dose and uniformity [20]. Positron dosimetry can nondestructively measure doses of implanted ions with significantly higher sensitivity than that available using other techniques. The principle of the technique is that implanted thermalized positrons diffusively move in the material and become trapped by the open-volume vacancy-type defects created by ion implantation. The positron annihilation in vacancy-type defects contributes less to the Doppler broadening of the energy spectrum of annihilation $\gamma$ rays compared to that in the defect-free bulk material. Doppler broadening parameter $S$ is defined as the ratio of the number of counts in the central part of 511 keV gamma line to the total number of counts under the peak. A single parameter $S$ describing the linewidth of the annihilation gamma ray line at 511 keV is related to the defect concentration. The concentration $C$ of open-volume defects is related to the number $\phi$ of ions implanted as $C \propto \phi^{0.7}$. The defect depth profiling using positron beams has found applications in materials research to study ion beam damage in both inorganic [21–23] and organic materials [24]. In the latter case, positron beam studies are of particular importance since application of X-ray or electron beams to organic materials may appear invasive [25].

Though modern MOS device technology may rely on ion-implantation free approaches [26, 27], applications of ion implantation are expanding over areas of quantum information processing [28, 29] and photovoltaics [30, 31]. Plasma immersion ion implantation enables fabrication of 3D transistor architectures [32, 33] required for scaling of metal-oxide-semiconductor field-effect transistors (MOSFETs) and is technologically more convenient for the fabrication of shallow pn-junctions. The ion implantation doping and the problems associated with the formation of point defects in the ion collision processes have been reviewed in work [34] highlighting the differences in the defect generation and accumulation in Si and Ge upon ion implantation. The dopant behavior in Ge is dominated by vacancies, while both vacancies and self-interstitials are active in Si. PAS has been applied to study point defects in interfaces between high-k dielectrics and metal [35] and Si [36]. The open volume defects were found to be located at both $TiN/SiO_2$ and $Si/SiO_2$ interfaces [37]. Annealing studies of defects indicated that while the defects in the $Si/SiO_2$ interface could be annealed out, the $TiN/SiO_2$ interface revealed an enhanced defect density due to the formation of the interfacial titanium oxynitride. Open volume defects introduced in $SiN_x$ films [38] and SiGe/Si interfaces [39] by plasma processing have been also revealed by PAS.

## 3. Electron spin resonance spectroscopy

Being integral to CMOS technology, ion implantation finds its applications at the forefront of materials science for fabrication of quasi-2D materials [40, 41], exploration of electron and nuclear spins of donor atoms in silicon as qubits for quantum information processing [42], and fabrication of light-emitting diodes [43]. Pertaining to MOS device fabrication, ion implantation is known to result in generation of electron and hole-trapping centers, which are detrimental to the device performance [44]. Such trapping centers may reside in a gate oxide and its interfaces with a semiconductor and a gate electrode. In amorphous $SiO_2$, ion implantation induces densification and the amorphous network reconstruction, not fully consistent with the assumption of plastic deformation. Ion implantation forces $SiO_2$ to freeze in a nonequilibrium phase tolerating a substantial reduction in the mean Si−O−Si angle and a subsequent change in the ring distribution statistics. As such, the radiation response of $SiO_2$ is dependent on the intrinsic structure of the material and the incorporated strain. Possible structural modifications in amorphous $SiO_2$ resulting in irradiation-induced charge have been reviewed in Ref. [45]. When paramagnetic, electrically active defects can be studied by using electron spin resonance (ESR) since the method is restricted to systems with a residual electron spin. For example, molecular solids with singlet ground states are not observable by ESR. This selectivity appears as useful in research on the electronic states of conducting materials, point defects in thin films, interfaces, and nanocrystals [46–50]. For the subject of ESR describing the fundamental theory and also the primary applications of the technique one can refer to the textbooks [51, 52]. The potential of the method in application to interfaces and nanolayers is detailed in Ref. [53].

The actual quantity detected in the ESR experiment is the net magnetic moment per unit volume, the macroscopic magnetization $M$. The microwave absorption spectrum is described by the spin Hamiltonian consisting of two components. A spin Hamiltonian contains operators for an effective electronic spin and for nuclear spins, the external magnetic field, and parameters. Its eigenfunctions determine the allowed energy levels of the system for an ESR experiment. The characteristics of paramagnetic species are the $g$-value, the spin-lattice relaxation time, and the line width. The $g$-value is the magnitude of the electron Zeeman factor for the paramagnetic species considered. The $g$-value can be determined as $E = g\mu_B B$, where $E$ is the energy of microwave, $\mu_B$ is Bohr magneton, and $B$ is magnetic field. In the case of free electrons, the $g$-value becomes 2.0023. For a paramagnetic defect, the $g$-value is different due to the effect of local magnetic field induced by movement of electrons in their orbits. The structure of the orbits contributes to the $g$-values via the effect of spin-orbit coupling, which is anisotropic and depends on axis determined by the magnetic field.

The spin-lattice relaxation time characterizes interactions of a spin system with its environment and reflects the strength of the interaction between the spin system and its surroundings. The magnetic environment of an unpaired electron can give rise to the ESR line broadening. The spectral lines are broadened either homogeneously or inhomogeneously. Homogeneous line broadening can be fitted by a single Lorentzian line and indicates that all the spins are described by the same spin Hamiltonian parameters. The line width of homogeneously broadened lines depends on the relaxation time of the spins. In the case of inhomogeneous broadening, the observed signal becomes a superposition of a large ensemble of individual spin packets, which are of slightly different $g$-values from each other. The inhomogeneous broadening of the spectral line can be caused, for example, by anisotropy of the $g$-tensor or the unresolved hyperfine structure. The latter may occur when the number of hyperfine components located near nuclei is so large that the hyperfine structure cannot be clearly observed. The large line width can be also observed due to dipole-dipole interactions between the defects spins [54].

As a starting point in defect identification, it is instructive to give an overview of intrinsic and extrinsic point defects of the $Si/SiO_2$ system as the most comprehensively studied system in CMOS technology. Being oxidized, silicon forms network-lattice-induced dangling bond defects at the $Si/SiO_2$ plane. The structure of the $P_b$ defects is dependent on the crystalline orientation of Si. The (111)$Si/SiO_2$ interface can be characterized by dangling bond defects of only one type—$P_b$ centers. This is a $sp3$ silicon-dangling bond directed along the [111]. The defect is of $C_{3v}$ symmetry and can exist in four orientations in the silicon lattice [55, 56]. Thermally oxidized silicon contains the $P_b$ density of approximately $4.9 \cdot 10^{12}$ см$^{-2}$. In contrast to the (111)$Si/SiO_2$ interface, the (100)$Si/SiO_2$ interface is characterized by two ESR active defects, $P_{b0}$ and $P_{b1}$ as shown in **Figure 3**. When oxidation of silicon is implemented at 800–970°C, the defect density of both defect types is similar ($10^{12}$ см$^{-2}$). The $P_{b1}$ defect is also a Si-dangling bond located slightly under the interface plane. Unlike $P_{b0}$, it is of monoclinic-I point symmetry [48].

The dangling bond silicon defects, the $P_b$ centers, are often employed as sensitive probes to detect interfacial stress during the $Si/SiO_2$ interface formation. When Si is subjected to oxidation at $T > 900°C$, structural relaxations occur at the $Si/SiO_2$, and the density of $P_b$-centers decreases. At this point, two stages of the silicon oxidation process can be distinguished. Suboxide bonding at the Si/SiO$_2$ interface is diminishing when silicon is oxidizing at 850°C<$T$< 900°C. Increasing oxidation temperature to 1050°C reduces strain at the macroscale [57]. Spatial uniformity of the dangling bond defects is determined by the temperature conditions during silicon oxidation. ESR studies of $P_b$ defects can be used to determine deformations at the interface from dependence of the ESR line width as a function of magnetic field angle [58, 59].

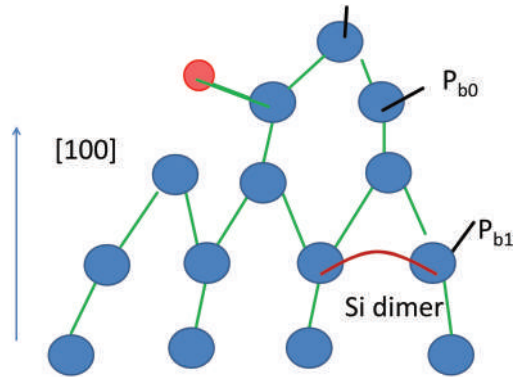

**Figure 3.** Schematic representation of $P_{b0}$ и $P_{b1}$ defects at the (100)Si/SiO$_2$ interface. After Ref. [145].

$P_{b0}$ and $P_{b1}$ defects in (100)Si/SiO$_2$ as well as $P_b$ defects in (111)Si/SiO$_2$ can be passivated in molecular hydrogen [49]. Upon ion implantation or ionizing irradiation, the interface trap generation may occur. A part of the interface states appears to be due to depassivated dangling bond defects. The mechanism of the depassivation reactions has been considered within the "hydrogen model", which assumes defect precursors in SiO$_2$ to create mobile protons interacting with $HP_b$ and generating $P_b$ centers. The interface trap generation coincides with the positive charge built-up in the oxide. The model proposes that protons are introduced in SiO$_2$ as a product of reactions of atomic hydrogen with the hole carriers trapped in the oxide; both the atomic hydrogen and the trapped holes are produced by irradiation. It has been concluded that the positive charge trapped in the oxide is present in the form of small polarons (self-trapped holes) in amorphous SiO$_2$ [60]. Though in bulk vitreous SiO$_2$ intrinsic hole-trap centers have been found to be stable at relatively low temperatures, thin films of insulating gate dielectrics in modern MOS devices are formed by low-temperature depositions on semiconductors and could incorporate interfacial strain sufficient to support self-trapped carriers at higher temperatures. The polaronic nature of the oxide-trapped charge in amorphous SiO$_2$ is consistent with the recent theoretical consideration of hole and electron trapping in hafnia. The deep states of electron and hole polarons have been predicted to exist in HfO$_2$ with precursor sites being elongated Hf−O bonds or under-coordinated Hf and O atoms [61]. This indicates that: (i) similar mechanisms of the defect generation under irradiation or ion beam damage could be operative in MOS devices containing HfO$_2$ and other amorphous oxides. (ii) Dangling bond defects in oxides may not be required for the charge trapping to occur.

Of the dangling bond defects in $SiO_2$, there are point defects associated with a dangling bond localized either on silicon or oxygen. The *EX* center belongs to the oxygen-related defects in $SiO_2$. The *EX* defect is the intrinsic network-stabilized defect in $SiO_2$. It is formed in the upper part of the oxide when the oxidation temperature $T_{ox}$ = 700–800°C. Being most prominent in thin oxides, *EX* is linked to the specific way thermal oxide is grown, i.e., oxidation of c-Si. As a working model, *EX* can be represented as an excess-O hole defect where an electron is delocalized over the four oxygen atoms bordering a Si vacancy [62], **Figure 4**. There are also a nonbridging oxygen hole center ($O_3{\equiv}Si{-}O\cdot$) [63] and a peroxide-radical ($Si{-}O{-}O\cdot$) [64], which are not naturally present in $SiO_2$ and introduced as damage defects in a postoxidation stage by irradiation with some energetic species (e.g., $\gamma$ and x photons, electrons, ions).

The *E′* defect is also an extrinsic defect present in crystalline and amorphous $SiO_2$. The *E′* defects in $SiO_2$ have an unpaired electron localized at a hybrid *sp3* orbital of silicon, which is bonded to three oxygen atoms ($O_3{\equiv}Si\cdot$) [65]. Several schematic models of the *E′* centers are depicted in **Figure 5**. The model representation of *E′* as the bridged hole-trapping oxygen-deficiency center has not been experientially verified [66]. The model considers a paramagnetic silicon atom connected via oxygen with another silicon atom, which is the trapping center for positive charge carriers, **Figure 5(b).** Generation of *E′* defects may depend on hydrogen content in a-$SiO_2$, since dissociation energy of a strained Si−O bond by hydrogen is rather low and amounts to 0.5–1.3 eB [67]. The defect generation in interfaces and thin films by ionizing radiation or hot electron injection is sensitive to the initial content of the strain bonds in MOS devices [68]. Therefore, ESR studies could be employed to reveal the impact of the interfacial strain on the defect generation.

Since electronic devices explore charge carries in their operation, it appeared natural to establish interrelationship between the silicon-dangling bond defects and the electron states at the semiconductor/insulator (SI) interfaces. For the Si/SiO$_2$ interfaces, it is known that technology chosen for silicon oxidation is crucial for attaining low density of the interface state ($D_{it}$), which is directly linked to the density of silicon-dangling bonds at the Si/SiO$_2$ interface. The decrease in $D_{it}$ and the $P_b$ density was observed when steam oxidation was used to grow SiO$_2$. Also, the higher $D_{it}$ values are expected at the more closely packed (111)Si surface as compared to the (100)Si one. A direct correlation between the $P_b$ density and the free carrier concentration in the field-effect transistor channel was reported in work [69]. Further studies of electrical activity of the Si/SiO$_2$ defects were undertaken by using various methodologies: capacitance-voltage (CV) measurements [70], deep-level transient spectroscopy [71], and the photoionization threshold method [72]. It was firmly established that $P_{b0}$ defects at the (100)Si/SiO$_2$ interface form amphoteric surface states at 0.3 and 0.8 eV above the silicon valence band edge [73]. In respect to the $P_{b1}$ centers at the (100)Si/SiO$_2$, the $P_{b0}$ и $P_{b1}$ defect densities inferred from ESR studies were compared with the interface trap densities determined from CV measurements. It was concluded that $P_{b1}$ does not form electrically active states within the silicon band gap [74]. Concerning the E′ center in thermal oxide, it is neutral when paramagnetic and strongly interacts with hydrogen [75]. The model for the E′ center in this case is the H-terminated center denoted as O$_3$≡Si–H. It has been supposed that the E′ center constitutes the hole trap and releases hydrogen in the form of a proton upon hole-trapping. The released proton can be trapped by the oxide network and form a donor-like surface state. When hydrogen is available in gate oxides as it can be upon an irradiation process, the neutral $E´$ center may be again passivated serving as a hole-trapping site.

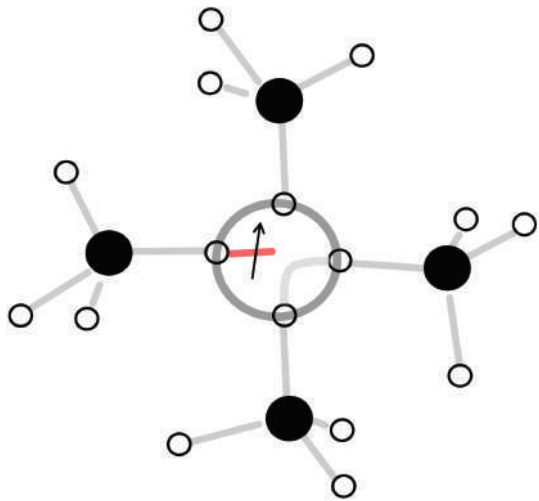

**Figure 4.** Schematic representation of the *EX* center.

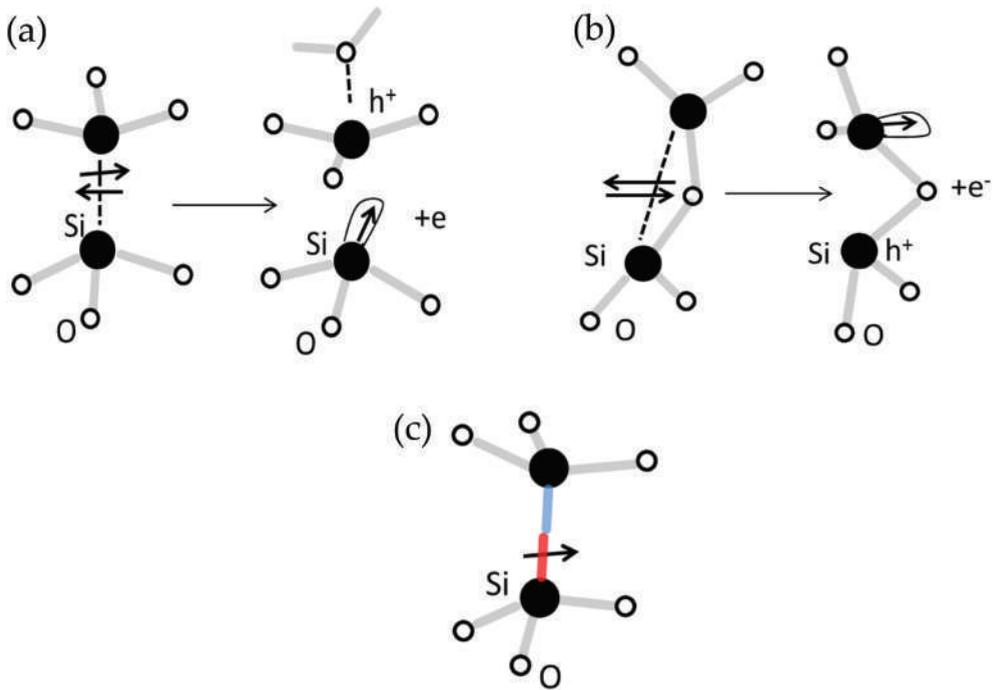

**Figure 5.** The first model of E′γ center (a), the model of the bridged hole-trapping oxygen-deficiency center (b), and the E′σ center model (c).

Charge trapping in gate oxides is one of the major obstacles in integration of high-k gate dielectrics in CMOS technology. Among the issues is the enhanced migration of dopant impurities originating from ion implantation steps. As such, ESR studies are indispensable to unravel point defects, which may appear detrimental for MOSFET performance. For example, ESR studies of phosphorous implanted high-k dielectrics reveal that P incorporating in the metal oxide network forms point defects by substituting for Hf or Zr in $HfO_2$ or $ZrO_2$, respectively [76]. Such defects formed due to enhanced migration of dopant impurities during dopant activation thermal steps may potentially trap charge. ESR studies have been applied to diverse ion-implanted systems. In $SiO_2$, a substantial reduction in $S$ and $E'_\gamma$ centers (Si enrichment in the oxide) was found when *in situ* ultrasound treatment was applied during implantation of $Si^+$ ions into thermal $SiO_2$ on (100)Si [77]; ESR found a radical mechanism of degradation of the ion-implanted photoresist [78]. Applications of the ESR techniques to study ion-beam-induced implantation damage in carbon-based materials have been described in Ref. [79].

ESR techniques have been explored in studies of spintronic materials fabricated by ion implantation. To probe the spin relaxation, the technique of choice is the pulse-electron spin resonance spectroscopy. ESR studies have been undertaken to measure spin relaxation times of dopants in Si. Shallow donors in Si are known for their long relaxation time suggesting a possible application of spins as qubits. The transverse relaxation time measured for isolated spins is associated with the decoherence time. ESR studies have been used to determine spin relaxation times in Sb-implanted isotopically enriched $^{28}$Si [80]. It has been shown that annealing of ultralow dose antimony implants leads to high degrees of electrical dopant activation with minimal diffusion. Spin relaxation times were increased when paramagnetic defects at the Si/$SiO_2$ interface were passivated by hydrogen. Except for the Si/$SiO_2$ system, pulsed ESR experiments have been used to characterize the coherent spin dynamics of nanofabricated nitrogen vacancy centers in nitrogen implanted high-purity diamond [81].

## 4. Electrical characterization of semiconductor interfaces: semiconductor doping, interfacial and oxide charges

### 4.1. Steady-state capacitance

An overview of charge carrier profiling, steady-state and transient capacitance, deep-level transient spectroscopy methods can be found in Ref. [82]. CV methods are most frequently used to extract parameters critical for operation of semiconductor devices. The interface trap densities, the fixed oxide charge, the carrier concentration in a semiconductor, and the permittivity of an insulator can be obtained from CV measurements. Here, more emphasis is given to basic limitations of CV methods, possible errors, and examples of using CV techniques.

The measure of charge responses in MOS devices as a function of electric field is the differential capacitance. To account for the interface trap effects, the Berglund method that establishes the relation between applied voltage across the MOS structure and the band-bending in equilibrium can be used [83]. **Figure 6** exemplifies the energy distributions of the interface states at the (100)Si/SiO$_2$ and (100)Si/HfO$_2$ interfaces in panels (a) and (b), respectively. The $D_{it}$ distributions in panel (a) are obtained using the Berglund procedure. The results of $D_{it}$ distributions extracted from a low-frequency CV curve and *ac* admittance data are compared in panel (b) indicating a perfect match in the energy range where the fast interface states contribute to the emission of charge carriers. The Si/SiO$_2$ interface trap distributions derived from the 100 Hz CV curves reveal two peaks centered at 0.25 and 0.85 eV above the Si valence-band edge, **Figure 6(a)**. The peaks are superimposed on the U-shaped background corresponding to a continuous distribution of the surface states in energy and ascribed to the existence of weak Si−Si and Si−O bonds at the Si/SiO$_2$interface [84]. The observed peak energy positions correspond to the (+/0) and (0/−) transitions of the amphoteric $P_{b0}$ defect. No measurable contribution of the $P_{b1}$ center to the $D_{it}$ could be detected in the central part of the Si bandgap, in agreement with the previous studies [74], which compare the total interface trap density $N_{it}$ and the $P_{b0}$ and $P_{b1}$ densities inferred from the ESR data. The estimation of the total interface trap density $N_{it}$ in work [74] was done according to Gray and Brown as described in work [85]. This procedure is advantageous over the low-frequency Berglund method in the following: (i) It allows detection of the interface states close (20 meV) to the Si band gap edges, inaccessible for room-temperature

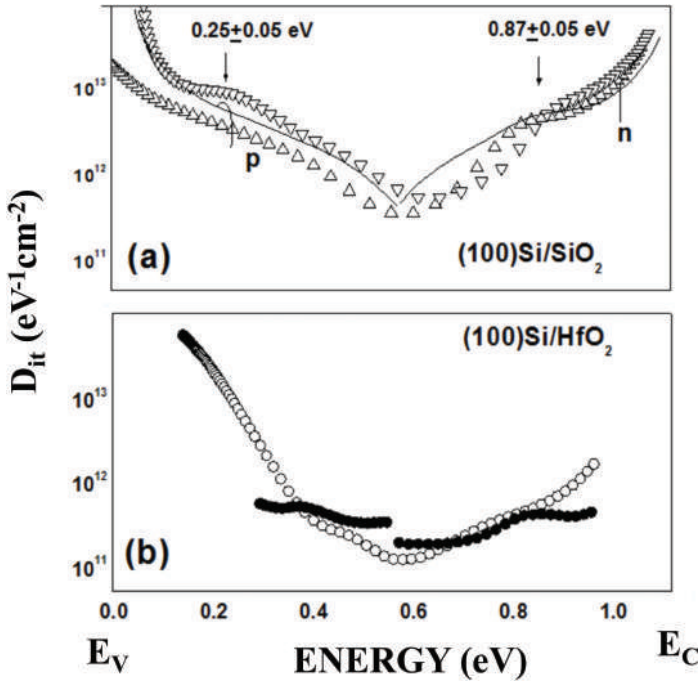

**Figure 6.** The interface trap distributions inferred from the capacitance-voltage measurements following the Berglund method and compared with these determined from *ac* conductance data as denoted by (•) symbols in panel (b).

CV analysis. (ii) It is decoupled from the uncertainty of Si surface potential determination near the band edges when the interface trap density is high.

A strong capacitance dispersion and dc leakage current may hamper application of CV methods based on low-high frequency CV measurements. As such, the Terman procedure based on comparison of the calculated ideal and experimental high-frequency CV curves may have limited applications in determining the interface trap densities in the case of interfaces of high-k dielectrics with semiconductors. Also, the interface trap contribution to the CV curve shift in voltage due to the interface traps becomes less prominent when equivalent electrical thickness of an insulator decreases [86, 87].

CV techniques can be used to extract the charge carrier profile in a semiconductor, the important characteristic of ion-implanted devices. The dopant profile is obtained from the high-frequency CV curve to minimize possible uncertainties stemming from the interface trap charge contribution to the depletion layer capacitance. The principle behind the dopant profiling is that as the semiconductor becomes depleted by the majority carriers, the capacitance decreases. A rapid decrease of the capacitance indicates a low dopant concentration, whereas a slow reduction indicates a high doping level. The capacitance as a function of voltage is related to the majority carrier density and can be obtained from the slope of the Mott-Schottky curve [88].

A variant of CV carrier profiling, which employs an electrochemical contact to a semiconductor, is an electrochemical capacitance-voltage (ECV) technique. ECV may appear as advantageous compared to the conventional CV methods due to its capability to measure spatial-ionized impurity distribution to practically unlimited depth, not being hindered by the breakdown at a high doping level [89]. ECV profiling can be applied to materials, which cannot be studied by the Hall measurements, for example, to conductive ferromagnetic semiconductors [90]. Despite its utility, ECV applicability is limited by the sample thickness when it is comparable with the Debye length, or, if a sample consists of several thin layers, which are either of different chemical composition or doping. The charge transfer at the interface is an important difference between a semiconductor/electrolyte (SE) interface and a Schottky contact. In the former case, it is supported by an electrochemical process. Parameters of the SE interface are determined by the electronic structure of the interface. The potential distribution in the SE interface and the effects of the semiconductor surface states on the potential redistribution between the semiconductor and the Helmholtz layer have been considered in review articles [91, 92]. When the surface states are not present at the semiconductor electrode the reverse bias drops across the semiconductor space charge region. It is than possible to determine the carrier concentration in the semiconductor. Except for the charge trapped in the surface states, there can be other charges, which result in the flat band voltage shift ($V_{fb}$) and modify the capacitive-frequency responses. An interfacial electric dipole layer can also result in a $V_{fb}$ shift when the latter coincides with the change in electron affinity indicating that the dipoles attached to the semiconductor surface contribute to the $V_{fb}$ shift, not surface charges.

Analysis of CMOS devices with nanometer thin insulators requires taking into account quantum-mechanical effects in the accumulation capacitance [93, 94] and the inversion capacitance [95] in order to extract the equivalent oxide thickness or the semiconductor doping density, respectively.

The doping density can be extracted from the inversion layer capacitance by relating the depletion layer width $W_D$ and the carrier concentration $N_{A,D}$

$$W_D = \sqrt{\frac{4\varepsilon_s \, kT ln(N_{A,D}/n_i)}{q \cdot (N_{A,D})}}, \qquad (3)$$

where $n_i$ is the intrinsic concentration in a semiconductor at a given temperature $T$ and $q$ is the elemental charge [96].

Alternatively, the doping density can be known from the band bending at the onset of strong inversion $\Psi_{s\,inv} \approx \frac{2kT}{q}\ln\left(\frac{N_{A,D}}{n_i}\right)$. The surface potential is obtained by using the Berglund integral. For the scaled MOS devices, one has to take into account the contribution of the finite density of states and the finite inversion layer thickness to the inversion layer capacitance or utilize the capacitance in the weak inversion to extract the substrate doping (cf. **Figure 2** in Ref. [95]). The inference of the semiconductor substrate doping from the inversion capacitance may appear to be superior over other experimental approaches, because it is decoupled from the possible contribution of the interface states to the depletion layer capacitance. This technique has been applied to trace boron concentration in silicon as a probe for the presence of radiolytic hydrogen in $SiO_2$ when analyzing the impact of vacuum ultraviolet irradiation and ion implantation of fluorine and argon on charge built-up in $Si/SiO_2$ MOS systems [97, 98]. Local characteristics of dopants can be obtained on semiconductor devices by using the scanning capacitance microscopy, a technique based on local capacitance-voltage analysis with submicron spatial resolution [99].

### 4.2. Steady-state *ac* conductance

The dynamic electrical responses of junction space-charge layers can be probed by using *ac* admittance spectroscopy or transient spectroscopy methods. These methods are applicable to both the deep bulk trap [100, 101] and interface trap [102, 103] studies in MOS devices. The *ac* admittance method is a classical approach to characterize the interface states in MOS structures [104]. The method better accentuates fast interface states, which are spatially located at the SI interface plane. The method considers the imaginary part of the measured admittance, which is directly linked to the charge trapped and emitted from the interface states as a consequence of the applied *ac* electric field. The localized states exchanging charge with the majority carrier band of a semiconductor respond to *ac* signal with both the capacitive and conductive components. At a particular frequency $\omega$ which is $\omega\tau = 1$, where $\tau$ is the characteristic time constant for the charge exchange with the localized state. The ratio $G_p/\omega$ reaches a maximum value directly proportional to the density of the surfaces states $D_{it}$. The trap occupancy is modulated by the semiconductor surface potential $\Psi_s$. The capture cross sections $\sigma_{p,n}$ and the trap densities $N_{t(p,n)}$ can be inferred from the frequency dependences of conductance exemplified in **Figure 7**. The interface trap resonances can be analyzed by using different models. Initially, it was suggested that there exists a quasicontinuous distribution of the interface states localized at the SI interface and that the surface charge and potential are uniform all over the interface. The broadening of the experimental normalized conductance curves was explained by Nicollian and Goetzberger as related to a random oxide charge and charge of the interface states distributed in the interface plane [105]. The tunnel recharging of the traps has been considered in Ref. [106]. To account for asymmetric conductance peaks, another model suggested that the interface traps at a particular energy have a range of cross sections spanning over orders of magnitude [107].

In nanoscale CMOS devices, the excessive leakage current impacts characterization of the interface traps by application of *ac* admittance spectroscopy. It has been demonstrated that errors in series resistance are critical when $D_{it}$ values are determined at the accumulation band bending, while high tunnel currents hamper characterization of the midgap interface states [108]. The practical solution of the problem associated with the interface trap characterization in tunnel MOS-devices is the use of the charge pumping method [109, 110]. When the leakage current does not impede the interface trap analysis, the interface states in the (100)Si/SiO$_2$ and (100)Si/HfO$_2$ entities can be reliably inferred from the capacitance frequency dispersion [111, 112] or *ac* admittance spectroscopy combined with the CV methods [113]. In the latter work, it has been observed that the $D_{it}$ density measured on Hf-containing samples subjected to a high-temperature anneal in oxygen and a subsequent passivation in hydrogen is still higher than that inferred for the equally treated (100)Si/SiO$_2$ interface. After passivation in molecular hydrogen, both the HfO$_2$ and SiO$_2$ interfaces with Si exhibited the $D_{it}$ peak positioned at 0.4 eV above the silicon valence band top. When $P_{b0}$ centers are passivated by molecular hydrogen the *ac* conductance responses are dominated by the contribution of the slow states, which are usually ascribed to the oxide-related imperfections. The slow states giving rise to the feature observed at 0.4 eV are likely to originate from the near interfacial oxide interlayer and could exist due to a lattice distortion in strained interfaces.

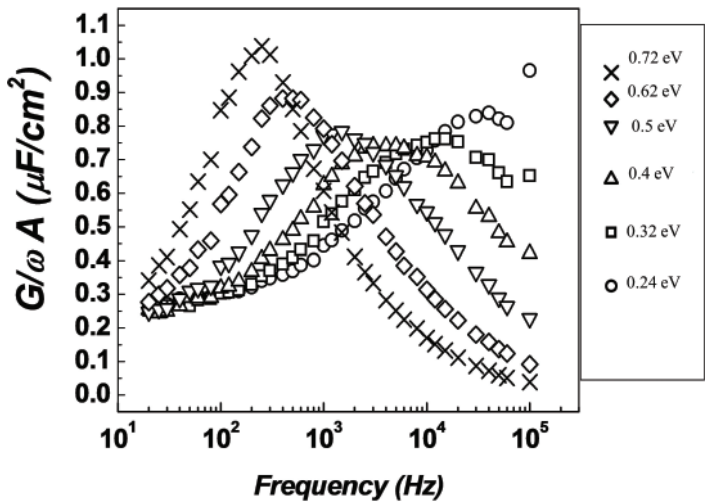

**Figure 7.** Equivalent parallel conductance as a function of frequency. The points are experimental values taken on a MOS-capacitor at different surface potentials $\psi_s$. The silicon substrate is of p-type conductivity.

## 4.3. Transient capacitance

Transient-capacitance spectroscopy has been initially developed to study deep bulk trap levels in semiconductors and termed by Lang as deep-level transient spectroscopy (DLTS). The capacitance DLTS is a preferred variant of the transient measurements, because it allows to separate minority and majority carrier emissions [114]. The technique is based on recording fast capacitance transients and passing the transient signal through a rate window circuit using a boxcar integrator and predefining the width of the gate pulse, the integrator response time, and the rate-window time constant. A lock-in amplifier used instead of a boxcar integrator requires settings for the rate-window, the initial gate-off period and the phase. When the traps are continuously distributed in energy (such as the interface traps) the measurement yields an emission time-constant spectrum, which depends on both the trap distribution and capture cross sections. A conventional DLTS procedure uses biases in depletion and pulsed voltage to populate interface traps with majority carriers. The responses of the device capacitance are recorded as the interface trap occupancy tends to equilibrium distribution. The energy of the traps can be determined independently of the emission rate by using two charging pulses of slightly different amplitude to selectively populate the interface traps [115]. A new method to determine capture cross sections independently of temperature and energy has been proposed in the work [116]. The method exploits the use of small trap-filling pulses to narrow the energy range within which the surface states become populated with majority carriers. Schematic diagrams representing (a) energy bands at the SI interface and (b) the pulsing sequence are shown in **Figure 8**.

When a voltage pulse sequence $\Delta V$ is superimposed on a constant voltage biasing a MOS structure to the surface depletion by the majority carriers, the capacitance difference recorded between times $t_1$ and $t_2$ is expressed as

$$\Delta C = A \int_{E_v}^{E_c} N_s(E) \left[ e^{-\frac{t_1}{\tau_n}} - e^{-\frac{t_2}{\tau_n}} \right] [f_0(E) - f_1(E)] dE, \qquad (4)$$

where $N_S(E)$ is the surface state density at energy $E$, $\tau_n$ is the emission time constant for electrons when considering n-type semiconductor. A constant $A = C_0^3/\varepsilon_s C_{ox} N_D$, where $C_0$ is a capacitance at reverse bias, $\varepsilon_s$ is the Si permittivity, and $C_{ox}$ is the insulator capacitance, $N_D$ is the substrate doping. The integration limits span from the valence band edge $E_v$ to the conductance band edge $E_c$, and $E_f$ is the Fermi level. The electron occupation of the surface states at the surface potential values $\Psi_s$ and $\Psi_s - \Delta E/q$ is described by the Fermi functions $f_o(E)$ and $f_1(E)$. As the pulse amplitude is small, the occupancy of the surface states can be approximated by the $\delta$ function, and Eq. (4) can be written as the capacitance of a discrete level.

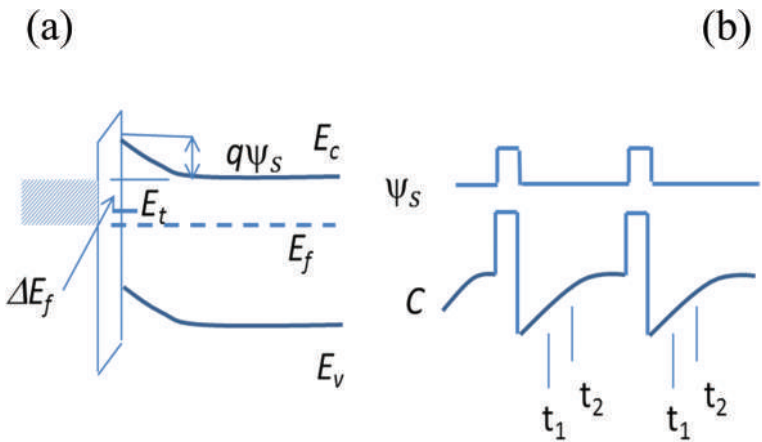

**Figure 8.** Schematic diagrams representing (a) energy bands at the SI interface and (b) the capacitance and the surface potential at the SI interface.

$$\Delta C = A N_s(E_t) \left[ e^{\left(-\frac{t_1}{\tau_n(E_t)}\right)} - e^{\left(-\frac{t_2}{\tau_n(E_t)}\right)} \right] \qquad (5)$$

For a discrete level, DLTS spectrum peaks at

$$\tau_n = \frac{t_2 - t_1}{\ln(t_2/t_1)} \qquad (6)$$

The emission time constant is expressed as

$$\tau_n = [v_{th} \cdot N_c \cdot \sigma_n e^{(-\Delta E_t/kT)}]^{-1}, \qquad (7)$$

where $v_{th}$ is the thermal velocity of electrons, $N_c = N_D e^{(qV_f/kT)}$ is the effective density of states in the conduction band, $\sigma_s$ is the capture cross section for electrons, and $\Delta E_t$ is the activation energy.

Assuming a capture cross section is exponentially dependent on energy

$$\sigma_n = \sigma_0 e^{-\Delta E_\sigma/kT}, \qquad (8)$$

with $\sigma_0$ and $\Delta E_\sigma$ being the preexponential factor and the activation energy, respectively, a set of the capture cross sections at different energies can be expressed as

$$\sigma_n(E_t, T) = \sigma_0(E_t) e^{\left(-\Delta E_\sigma(E_t)/kT\right)}. \qquad (9)$$

The apparent activation energy and the energy-dependent term $\sigma_0(E_t)$ can be determined from the Arrhenius plot. Repeating the DLTS measurements at different gate voltages (i.e., different surface potentials), one obtains $\sigma_0(E_t)$. The surface potential values can be determined from CV curves. The doping density and the oxide capacitance are estimated from the CV curves under the inversion and the accumulation, respectively.

In DLTS measurements, the bias dependence of the peak is a distinct signature of the charge carrier emission from the interface states [117]. Being characterized by DLTS and CV measurements, the oxide charge, the interface state densities, and capture cross sections in the energy gap can be utilized to obtain surface recombination velocities [118]. Applying DLTS pulses of opposite polarity (from accumulation to inversion) allows estimating the thermal generation times of bulk and surface centers [119]. DLTS techniques are capable in determining the trap properties in terms of relaxation mechanism and the defect profiling, the information valuable to study defects introduced by ion beams and ionizing radiation [100]. Naturally, characterization approaches are purpose-specific and can be based on several experimental techniques to identify a particular defect or study its energetics and kinetics. For example, commonly used techniques for studying the electrical- and optical characteristics of point defects such as DLTS and photoluminescence are sensitive to the defect states within the bandgap but have to be complemented by ESR studies to obtain information on the atomic structure of a defect or a defect complex.

**4.4. Photoinjection**

The methods based on photoinjection of charge carriers in metal-semiconductor barrier structures are sensitive to local nonuniformities in semiconductor interfaces because charge in a semiconductor induces an equal charge in the electrodes giving rise to electric fields at the interfaces, with a consequent field-effect modulation of the barrier heights (for the all-encompassing review on the subject of internal photoemission spectroscopy (IPE) methods one can refer to the book [120]). The early application of scanning internal photoemission to map sodium contamination at the $Si/SiO_2$ interface has been reported in work [121]. The IPE and trap photodepopulation methods were applied to reveal electron traps in $Na^+$ and $Al^+$ implanted $SiO_2$ [122]. At present, this technique has been revived to study ion beam induced charge nonuniformities in GaN and SiC [123].

Experimentally, the charge injected into an oxide, i.e., the current over the time of injection should remain unchanged by the method used for the charge detection. The trapped charge density is determined sensing the electric field created by the trapped charge. The electric field created by the charge of trapped carriers can be also observed in variations of the surface band bending of a semiconductor, i.e., a semiconductor space-charge layer serves as the field-sensing element. The band bending as a function of electric field can be extracted from capacitance-voltage measurements and the additional contribution of trapped charge to the field can be determined as a voltage shift of a CV curve. In MOSFETs, the trapped charge can be monitored as a function of the threshold voltage. This technique senses the charge carrier density in the inversion channel to monitor the electric field at the SI interface. Alternatively, the electric field induced by the trapped charge can be monitored by the Kelvin probe or photovoltage. In the latter case, the light intensity should be sufficient to set the flat band conditions at the semiconductor surface.

The experimental studies of the trapped charge in ion-implanted insulators are numerous with several examples represented in Refs. [124–131]. The interfacial defect densities modified by ion implantation have been studied combining IPE and *ac* conductance spectroscopy methods on nitrogen implanted SiC/SiO$_2$ interfaces [132]. IPE reveals that nitrogen incorporates in carbon clusters at the SiC/SiO$_2$ interface that causes a shift of the electron levels to higher binding energies. Inferring the Schottky barrier height from the IPE spectra, it has been shown that ion implantation of sulfur in the NiSi/Si barrier does not induce changes in the barrier height, but increases doping in silicon [133]. The silicide/Si barrier modification by intentional dopant segregation has been verified in work [134].

**4.5. Slow interface states as a special case of study**

Defects generated by ionizing radiation and/or electric field, as well as the defects in undamaged devices, are considered to be spatially distributed across the SI interface and can be classified accordingly to the spatial location as the oxide-related traps and the interface traps. In respect to the latter, it is generally accepted that the interface traps are rapidly communicating with the silicon conduction or valence bands. The defects within the oxide interlayer also can exchange charge with silicon as has been revealed by the noise measurements [135]. Combining *ac* admittance spectroscopy and the noise measurements, it has been established that the fast interface states at the Si/SiO$_2$ interface, likely associated with the dangling bond defects, contribute to the loss peak in conductance measurements [136]. The defect states residing in an oxide layer are responsible for 1/f noise and random telegraph noise. These trapping centers in the oxide contribute to the conductance plateau at low frequency in *ac* conductance spectra (cf. **Figure 2** of Ref. [137]). A separable contribution of the oxide-related traps has been revealed employing measurements of subthreshold current [138] and the charge-pumping technique [139] to MOSFETs and CV measurements taken on the gate-controlled diode [140]. The latter technique is applicable for characterization of the interface traps in MOS devices composed on wide band gap semiconductors, because it allows supplying minority carriers in an amount sufficient to compensate for the low thermal generation rates of the minority carriers. An alternative method of providing minority carriers to invert a semiconductor surface is a controlled deposition of surface charges onto an insulator surface from corona discharging in air as it has been proposed in Ref. [141]. In this work, a surface charge has been deposited on SiO$_2$ and high-k dielectrics to overcompensate the carrier leakage current in silicon MOS capacitors and enable extraction of $D_{it}(E)$ profiles following the Berglund formalism. There are several advantages of the inverting semiconductor surfaces by employing noninvasive electrostatic charging of an insulator surface in a MOS structure: (i) The method does not involve fabrication of a transistor or a gate-controlled diode. (ii) The Berglund analysis can be used to reliably estimate $D_{it}(E)$ over the major part of a semiconductor band gap (for Si, from 0.2 to 0.9 eV above the valence band edge) using just MOS capacitors of one type of semiconductor conductivity. (iii) The method may employ CV measurements at mid-kHz frequency range allowing investigation of samples, which experience relatively high leakage current.

The sub-division of the interface trap responses into slow and fast on the basis of their characteristic time constants is important in research on the irradiation-induced damage in MOS devices. The interface state generation under irradiation or high electric field stress can involve electron-hole recombination in a gate insulator as proposed by Lai [142], the hole trapping according to the "hydrogen model" by Griscom [60], or generation of dangling bond defects in the oxide. Experimentally, it has been shown that both the fast and slow interface states can be generated upon oxide damage by high electric field or irradiation [143]. The mechanisms operative in the interface trap built-up upon irradiation or electric field stress are governed by hydrogen impurity, interfacial strain preexisting in thin insulating films on semiconductors, and experimental conditions used to impose damage on MOS devices.

## Author details

Yanina G. Fedorenko

Address all correspondence to: janina.fedorenko@gmail.com

Stephenson Institute for Renewable Energy and Department of Physics, School of Physical Sciences, Chadwick Building, University of Liverpool, Liverpool, UK